
\input phyzzx
\catcode`\@=11
\paperfootline={\hss\iffrontpage\else\ifp@genum\tenrm
 -- \folio\ --\hss\fi\fi}
\def\titlestyle#1{\par\begingroup \titleparagraphs
 \iftwelv@\fourteenpoint\fourteenbf\else\twelvepoint\twelvebf\fi
 \noindent #1\par\endgroup }
\def\GENITEM#1;#2{\par \hangafter=0 \hangindent=#1
 \Textindent{#2}\ignorespaces}
\def\address#1{\par\kern 5pt\titlestyle{\twelvepoint\sl #1}}
\def\abstract{\par\dimen@=\prevdepth \hrule height\z@ \prevdepth=\dimen@
 \vskip\frontpageskip\centerline{\fourteencp Abstract}\vskip\headskip }
\newif\ifYUKAWA  \YUKAWAtrue
\font\elevenmib   =cmmib10 scaled\magstephalf   \skewchar\elevenmib='177
\def\YUKAWAmark{\hbox{\elevenmib
 Yukawa\hskip0.05cm Institute\hskip0.05cm Kyoto \hfill}}
\def\titlepage{\FRONTPAGE\papers\ifPhysRev\PH@SR@V\fi
 \ifYUKAWA\null\vskip-1.70cm\YUKAWAmark\vskip0.6cm\fi
 \ifp@bblock\p@bblock \else\hrule height\z@ \rel@x \fi }

\def\schapter#1{\par \penalty-300 \vskip\chapterskip
 \spacecheck\chapterminspace
 \chapterreset \titlestyle{\ifcn@@\S\ \chapterlabel.~\fi #1}
 \nobreak\vskip\headskip \penalty 30000
 {\pr@tect\wlog{\string\chapter\space \chapterlabel}} }

\def\ssection#1{\par \ifnum\lastpenalty=30000\else
 \penalty-200\vskip\sectionskip \spacecheck\sectionminspace\fi
 \gl@bal\advance\sectionnumber by 1
 {\pr@tect
 \xdef\sectionlabel{\ifcn@@ \chapterlabel.\fi
 \the\sectionstyle{\the\sectionnumber}}%
 \wlog{\string\section\space \sectionlabel}}%
 \noindent {\S \caps\thinspace\sectionlabel.~~#1}\par
 \nobreak\vskip\headskip \penalty 30000 }


\papers

\def\lkakko{\vbox{\vskip0.065cm\hbox{(}\vskip-0.065cm}}
\def\rkakko{\vbox{\vskip0.065cm\hbox{)}\vskip-0.065cm}}
\def\YUKAWAHALL{\hbox to \hsize
 {\hfil \lkakko\twelvebf YUKAWA HALL\rkakko\hfil}}


\def\Endline{\hfil\break}


\def\addeqno{\ifnum\equanumber<0 \global\advance\equanumber by -1
 \else \global\advance\equanumber by 1\fi}


\mathchardef\Lag="724C
\def\sqr#1#2{{\vcenter{\hrule height.#2pt
 \hbox{\vrule width.#2pt height#1pt \kern#1pt\vrule width.#2pt}
 \hrule height.#2pt}}}

\def\square{{\mathchoice{\sqr84}{\sqr84}{\sqr{5.0}3}{\sqr{3.5}3}}}

\def\cref#1{\rlap,\attach{#1)}}
\def\ref#1{\attach{#1)}}

\def\ssatop#1#2{{\displaystyle\mathop{#2}^{\scriptscriptstyle #1}}}


\newdimen\ex@
\ex@.2326ex
\def\boxed#1{\setbox\z@\hbox{$\displaystyle{#1}$}\hbox{\lower.4\ex@
 \hbox{\lower3\ex@\hbox{\lower\dp\z@\hbox{\vbox{\hrule height.4\ex@
 \hbox{\vrule width.4\ex@\hskip3\ex@\vbox{\vskip3\ex@\box\z@\vskip3\ex@}%
 \hskip3\ex@\vrule width.4\ex@}\hrule height.4\ex@}}}}}}
\def\txtboxed#1{\setbox\z@\hbox{{#1}}\hbox{\lower.4\ex@
 \hbox{\lower3\ex@\hbox{\lower\dp\z@\hbox{\vbox{\hrule height.4\ex@
 \hbox{\vrule width.4\ex@\hskip3\ex@\vbox{\vskip3\ex@\box\z@\vskip3\ex@}%
 \hskip3\ex@\vrule width.4\ex@}\hrule height.4\ex@}}}}}}
\newdimen\exx@
\exx@.1ex
\def\thinboxed#1{\setbox\z@\hbox{$\displaystyle{#1}$}\hbox{\lower.4\exx@
 \hbox{\lower3\exx@\hbox{\lower\dp\z@\hbox{\vbox{\hrule height.4\exx@
 \hbox{\vrule width.4\exx@\hskip3\exx@%
 \vbox{\vskip3\ex@\box\z@\vskip3\exx@}%
 \hskip3\exx@\vrule width.4\exx@}\hrule height.4\exx@}}}}}}

\chardef\fontD="1A

\catcode`@=12

%
\catcode`\@=11
\def\relaxnext@{\let\next\relax}
\def\hexnumber@#1{\ifcase#1 0\or1\or2\or3\or4\or5\or6\or7\or8\or9\or
 A\or B\or C\or D\or E\or F\fi}
\edef\bffam@{\hexnumber@\bffam}
\mathchardef\boldGamma="0\bffam@00
\mathchardef\boldDelta="0\bffam@01
\mathchardef\boldTheta="0\bffam@02
\mathchardef\boldLambda="0\bffam@03
\mathchardef\boldXi="0\bffam@04
\mathchardef\boldPi="0\bffam@05
\mathchardef\boldSigma="0\bffam@06
\mathchardef\boldUpsilon="0\bffam@07
\mathchardef\boldPhi="0\bffam@08
\mathchardef\boldPsi="0\bffam@09
\mathchardef\boldOmega="0\bffam@0A
\font\tenmsx=msam10
\font\sevenmsx=msam7
\font\fivemsx=msam5
\font\tenmsy=msbm10
\font\sevenmsy=msbm7
\font\fivemsy=msbm5
\newfam\msxfam
\newfam\msyfam
\textfont\msxfam=\tenmsx
\scriptfont\msxfam=\sevenmsx
\scriptscriptfont\msxfam=\fivemsx
\textfont\msyfam=\tenmsy
\scriptfont\msyfam=\sevenmsy
\scriptscriptfont\msyfam=\fivemsy
\font\teneuf=eufm10
\font\seveneuf=eufm7
\font\fiveeuf=eufm5
\newfam\euffam
\textfont\euffam=\teneuf
\scriptfont\euffam=\seveneuf
\scriptscriptfont\euffam=\fiveeuf
\def\frak{\relaxnext@\ifmmode\let\next\frak@\else
 \def\next{\Err@{Use \string\frak\space only in math mode}}\fi\next}
\def\goth{\relaxnext@\ifmmode\let\next\frak@\else
 \def\next{\Err@{Use \string\goth\space only in math mode}}\fi\next}
\def\frak@#1{{\frak@@{#1}}}
\def\frak@@#1{\noaccents@\fam\euffam#1}

%
\edef\msx@{\hexnumber@\msxfam}
\edef\msy@{\hexnumber@\msyfam}
\mathchardef\boxdot="2\msx@00
\mathchardef\boxplus="2\msx@01
\mathchardef\boxtimes="2\msx@02
\mathchardef\square="0\msx@03
\mathchardef\blacksquare="0\msx@04
\mathchardef\centerdot="2\msx@05
\mathchardef\lozenge="0\msx@06
\mathchardef\blacklozenge="0\msx@07
\mathchardef\circlearrowright="3\msx@08
\mathchardef\circlearrowleft="3\msx@09
\mathchardef\leftrightharpoons="3\msx@0B
\mathchardef\boxminus="2\msx@0C
\mathchardef\Vdash="3\msx@0D
\mathchardef\Vvdash="3\msx@0E
\mathchardef\vDash="3\msx@0F
\mathchardef\twoheadrightarrow="3\msx@10
\mathchardef\twoheadleftarrow="3\msx@11
\mathchardef\leftleftarrows="3\msx@12
\mathchardef\rightrightarrows="3\msx@13
\mathchardef\upuparrows="3\msx@14
\mathchardef\downdownarrows="3\msx@15
\mathchardef\upharpoonright="3\msx@16

\mathchardef\downharpoonright="3\msx@17
\mathchardef\upharpoonleft="3\msx@18
\mathchardef\downharpoonleft="3\msx@19
\mathchardef\rightarrowtail="3\msx@1A
\mathchardef\leftarrowtail="3\msx@1B
\mathchardef\leftrightarrows="3\msx@1C
\mathchardef\rightleftarrows="3\msx@1D
\mathchardef\Lsh="3\msx@1E
\mathchardef\Rsh="3\msx@1F
\mathchardef\rightsquigarrow="3\msx@20
\mathchardef\leftrightsquigarrow="3\msx@21
\mathchardef\looparrowleft="3\msx@22
\mathchardef\looparrowright="3\msx@23
\mathchardef\circeq="3\msx@24
\mathchardef\succsim="3\msx@25
\mathchardef\gtrsim="3\msx@26
\mathchardef\gtrapprox="3\msx@27
\mathchardef\multimap="3\msx@28
\mathchardef\therefore="3\msx@29
\mathchardef\because="3\msx@2A
\mathchardef\doteqdot="3\msx@2B

\mathchardef\triangleq="3\msx@2C
\mathchardef\precsim="3\msx@2D
\mathchardef\lesssim="3\msx@2E
\mathchardef\lessapprox="3\msx@2F
\mathchardef\eqslantless="3\msx@30
\mathchardef\eqslantgtr="3\msx@31
\mathchardef\curlyeqprec="3\msx@32
\mathchardef\curlyeqsucc="3\msx@33
\mathchardef\preccurlyeq="3\msx@34
\mathchardef\leqq="3\msx@35
\mathchardef\leqslant="3\msx@36
\mathchardef\lessgtr="3\msx@37
\mathchardef\backprime="0\msx@38
\mathchardef\risingdotseq="3\msx@3A
\mathchardef\fallingdotseq="3\msx@3B
\mathchardef\succcurlyeq="3\msx@3C
\mathchardef\geqq="3\msx@3D
\mathchardef\geqslant="3\msx@3E
\mathchardef\gtrless="3\msx@3F
\mathchardef\sqsubset="3\msx@40
\mathchardef\sqsupset="3\msx@41
\mathchardef\vartriangleright="3\msx@42
\mathchardef\vartriangleleft ="3\msx@43
\mathchardef\trianglerighteq="3\msx@44
\mathchardef\trianglelefteq="3\msx@45
\mathchardef\bigstar="0\msx@46
\mathchardef\between="3\msx@47
\mathchardef\blacktriangledown="0\msx@48
\mathchardef\blacktriangleright="3\msx@49
\mathchardef\blacktriangleleft="3\msx@4A
\mathchardef\vartriangle="0\msx@4D
\mathchardef\blacktriangle="0\msx@4E
\mathchardef\triangledown="0\msx@4F
\mathchardef\eqcirc="3\msx@50
\mathchardef\lesseqgtr="3\msx@51
\mathchardef\gtreqless="3\msx@52
\mathchardef\lesseqqgtr="3\msx@53
\mathchardef\gtreqqless="3\msx@54
\mathchardef\Rrightarrow="3\msx@56
\mathchardef\Lleftarrow="3\msx@57
\mathchardef\veebar="2\msx@59
\mathchardef\barwedge="2\msx@5A
\mathchardef\doublebarwedge="2\msx@5B
\mathchardef\measuredangle="0\msx@5D
\mathchardef\sphericalangle="0\msx@5E
\mathchardef\varpropto="3\msx@5F
\mathchardef\smallsmile="3\msx@60
\mathchardef\smallfrown="3\msx@61
\mathchardef\Subset="3\msx@62
\mathchardef\Supset="3\msx@63
\mathchardef\Cup="2\msx@64

\mathchardef\Cap="2\msx@65

\mathchardef\curlywedge="2\msx@66
\mathchardef\curlyvee="2\msx@67
\mathchardef\leftthreetimes="2\msx@68
\mathchardef\rightthreetimes="2\msx@69
\mathchardef\subseteqq="3\msx@6A
\mathchardef\supseteqq="3\msx@6B
\mathchardef\bumpeq="3\msx@6C
\mathchardef\Bumpeq="3\msx@6D
\mathchardef\lll="3\msx@6E

\mathchardef\ggg="3\msx@6F

\mathchardef\circledS="0\msx@73
\mathchardef\pitchfork="3\msx@74
\mathchardef\dotplus="2\msx@75
\mathchardef\backsim="3\msx@76
\mathchardef\backsimeq="3\msx@77
\mathchardef\complement="0\msx@7B
\mathchardef\intercal="2\msx@7C
\mathchardef\circledcirc="2\msx@7D
\mathchardef\circledast="2\msx@7E
\mathchardef\circleddash="2\msx@7F
\def\ulcorner{\delimiter"4\msx@70\msx@70 }
\def\urcorner{\delimiter"5\msx@71\msx@71 }
\def\llcorner{\delimiter"4\msx@78\msx@78 }
\def\lrcorner{\delimiter"5\msx@79\msx@79 }
\mathchardef\lvertneqq="3\msy@00
\mathchardef\gvertneqq="3\msy@01
\mathchardef\nleq="3\msy@02
\mathchardef\ngeq="3\msy@03
\mathchardef\nless="3\msy@04
\mathchardef\ngtr="3\msy@05
\mathchardef\nprec="3\msy@06
\mathchardef\nsucc="3\msy@07
\mathchardef\lneqq="3\msy@08
\mathchardef\gneqq="3\msy@09
\mathchardef\nleqslant="3\msy@0A
\mathchardef\ngeqslant="3\msy@0B
\mathchardef\lneq="3\msy@0C
\mathchardef\gneq="3\msy@0D
\mathchardef\npreceq="3\msy@0E
\mathchardef\nsucceq="3\msy@0F
\mathchardef\precnsim="3\msy@10
\mathchardef\succnsim="3\msy@11
\mathchardef\lnsim="3\msy@12
\mathchardef\gnsim="3\msy@13
\mathchardef\nleqq="3\msy@14
\mathchardef\ngeqq="3\msy@15
\mathchardef\precneqq="3\msy@16
\mathchardef\succneqq="3\msy@17
\mathchardef\precnapprox="3\msy@18
\mathchardef\succnapprox="3\msy@19
\mathchardef\lnapprox="3\msy@1A
\mathchardef\gnapprox="3\msy@1B
\mathchardef\nsim="3\msy@1C
\mathchardef\ncong="3\msy@1D

\mathchardef\varsubsetneq="3\msy@20
\mathchardef\varsupsetneq="3\msy@21
\mathchardef\nsubseteqq="3\msy@22
\mathchardef\nsupseteqq="3\msy@23
\mathchardef\subsetneqq="3\msy@24
\mathchardef\supsetneqq="3\msy@25
\mathchardef\varsubsetneqq="3\msy@26
\mathchardef\varsupsetneqq="3\msy@27
\mathchardef\subsetneq="3\msy@28
\mathchardef\supsetneq="3\msy@29
\mathchardef\nsubseteq="3\msy@2A
\mathchardef\nsupseteq="3\msy@2B
\mathchardef\nparallel="3\msy@2C
\mathchardef\nmid="3\msy@2D
\mathchardef\nshortmid="3\msy@2E
\mathchardef\nshortparallel="3\msy@2F
\mathchardef\nvdash="3\msy@30
\mathchardef\nVdash="3\msy@31
\mathchardef\nvDash="3\msy@32
\mathchardef\nVDash="3\msy@33
\mathchardef\ntrianglerighteq="3\msy@34
\mathchardef\ntrianglelefteq="3\msy@35
\mathchardef\ntriangleleft="3\msy@36
\mathchardef\ntriangleright="3\msy@37
\mathchardef\nleftarrow="3\msy@38
\mathchardef\nrightarrow="3\msy@39
\mathchardef\nLeftarrow="3\msy@3A
\mathchardef\nRightarrow="3\msy@3B
\mathchardef\nLeftrightarrow="3\msy@3C
\mathchardef\nleftrightarrow="3\msy@3D
\mathchardef\divideontimes="2\msy@3E
\mathchardef\varnothing="0\msy@3F
\mathchardef\nexists="0\msy@40
\mathchardef\mho="0\msy@66
\mathchardef\eth="0\msy@67
\mathchardef\eqsim="3\msy@68
\mathchardef\beth="0\msy@69
\mathchardef\gimel="0\msy@6A
\mathchardef\daleth="0\msy@6B
\mathchardef\lessdot="3\msy@6C
\mathchardef\gtrdot="3\msy@6D
\mathchardef\ltimes="2\msy@6E
\mathchardef\rtimes="2\msy@6F
\mathchardef\shortmid="3\msy@70
\mathchardef\shortparallel="3\msy@71
\mathchardef\smallsetminus="2\msy@72
\mathchardef\thicksim="3\msy@73
\mathchardef\thickapprox="3\msy@74
\mathchardef\approxeq="3\msy@75
\mathchardef\succapprox="3\msy@76
\mathchardef\precapprox="3\msy@77
\mathchardef\curvearrowleft="3\msy@78
\mathchardef\curvearrowright="3\msy@79
\mathchardef\digamma="0\msy@7A
\mathchardef\varkappa="0\msy@7B
\mathchardef\hslash="0\msy@7D
\mathchardef\backepsilon="3\msy@7F
\def\accentfam@{7}
\def\noaccents@{\def\accentfam@{0}}
\catcode`\@=\active


\catcode`\@=11
\font\fourteenmib =cmmib10 scaled\magstep2  \skewchar\fourteenmib='177
\font\twelvemib   =cmmib10 scaled\magstep1   \skewchar\twelvemib='177
\font\tenmib      =cmmib10  \skewchar\tenmib='177
\font\fourteenbsy  =cmbsy10 scaled\magstep2  \skewchar\fourteenbsy='60
\font\twelvebsy    =cmbsy10 scaled\magstep1	  \skewchar\twelvebsy='60
\font\elevenbsy    =cmbsy10 scaled\magstephalf  \skewchar\elevenbsy='60
\font\tenbsy       =cmbsy10    \skewchar\tenbsy='60
\newfam\mibfam
\def\mib{\n@expand\f@m\mibfam}
\let\tmpfourteenf@nts=\fourteenf@nts
\def\fourteenf@nts{\tmpfourteenf@nts %
 \textfont\mibfam=\fourteenmib \scriptfont\mibfam=\tenmib
 \scriptscriptfont\mibfam=\tenmib }
\let\tmptwelvef@nts=\twelvef@nts
\def\twelvef@nts{\tmptwelvef@nts %
 \textfont\mibfam=\twelvemib \scriptfont\mibfam=\tenmib
 \scriptscriptfont\mibfam=\tenmib }
\let\tmptenf@nts=\tenf@nts
\def\tenf@nts{\tmptenf@nts %
    \textfont\mibfam=\tenmib   \scriptfont\mibfam=\tenmib
    \scriptscriptfont\mibfam=\tenmib }
\mathchardef\alpha     ="710B
\mathchardef\beta      ="710C
\mathchardef\gamma     ="710D
\mathchardef\delta     ="710E
\mathchardef\epsilon   ="710F
\mathchardef\zeta      ="7110
\mathchardef\eta       ="7111
\mathchardef\theta     ="7112
\mathchardef\iota      ="7113
\mathchardef\kappa     ="7114
\mathchardef\lambda    ="7115
\mathchardef\mu        ="7116
\mathchardef\nu        ="7117
\mathchardef\xi        ="7118
\mathchardef\pi        ="7119
\mathchardef\rho       ="711A
\mathchardef\sigma     ="711B
\mathchardef\tau       ="711C
\mathchardef\upsilon   ="711D
\mathchardef\phi       ="711E
\mathchardef\chi       ="711F
\mathchardef\psi       ="7120
\mathchardef\omega     ="7121
\mathchardef\varepsilon="7122
\mathchardef\vartheta  ="7123
\mathchardef\varpi     ="7124
\mathchardef\varrho    ="7125
\mathchardef\varsigma  ="7126
\mathchardef\varphi    ="7127

\def\fourteenpoint{\fourteenf@nts \samef@nt \b@gheight=14pt \setstr@t }
\def\twelvepoint{\twelvef@nts \samef@nt \b@gheight=12pt \setstr@t }
\def\tenpoint{\tenf@nts \samef@nt \b@gheight=10pt \setstr@t }
\def\Tenpoint{\tenpoint\twelv@false\spaces@t}
\def\Twelvepoint{\twelvepoint\twelv@true\spaces@t}
\Twelvepoint  
\catcode`\@=12

\def\overset#1\to#2{\mathop{#2}\limits^{#1}}
\def\overset#1\to#2{\ssatop{#2}\limits^{#1}}
\def\rhup{\rightharpoonup}
\pubnum={YITP/K-1069}
\date={April 1994}

\titlepage
\title{
SURFACE MOTIONS and FLUID DYNAMICS\footnote*{%
Talk given at the workshop on `Membranes and Higher Dimensional
Extended
Objects'\Endline
( March 1994, Isaac Newton Institute for Mathematical Sciences)}
}

\author{Jens Hoppe\footnote{**}{%
Heisenberg Fellow\Endline
On leave of absence from the Institute for Theoretical Physics,
Karlsruhe University.}}
\address{
Yukawa Institute for Theoretical Physics\break
Kyoto University,~Kyoto 606,~Japan}

\abstract{
A certain class of surface motions, including those of a relativistic
membrane minimizing the 3-dimensional volume swept out
in Minkowski-space, is shown to be equivalent to 3-dimensional
steady-state irrotational inviscid isentropic gas-dynamics.
The SU($\infty $) Nahm equations turn out to correspond to
motions where the time $t$ at which the surface moves
through the point $\ssatop\rhup{r}$ is a harmonic function of
the three space-coordinates.
This solution also implies the linearisation of a non-trivial-looking
scalar field theory.
}

\endpage

The dynamics of surfaces is of vital interest to quite a variety of
physical problems (see e.g. [1]).
Models that have attracted particular attention include the evolution
by mean curvature ([2], [3], [4], [5]), the kinetic roughening of
growing surfaces and the motion of domain walls [14], developable
surfaces and surfaces maintaining constant negative curvature
(see e.g. [6], [7]), and in [8], several families of integrable
surfaces have been determined, for which the
spectral parameter appearing in the zero-curvature condition may be
interpreted as a (time-) deformation parameter.

Let me consider a time-dependent 2-dimensional surface $\Sigma _t$ in
${\bf R}^3$,
whose motion is locally of the form
$$
\dot{\mathop{r}^\rightharpoonup} = a \cdot
\mathop{n}^\rightharpoonup~~,
\eqno{(1)}
$$
where $\ssatop{\rhup}{r} =
\ssatop\rhup{r}~(t,~\varphi^1,~\varphi^2)$ is the position vector,
$\ssatop{\rhup}{n}$ the surface normal, and $a = a(g)$ is assumed to be
only a function of the surface area element
$\sqrt g =
\mid \partial_1 \ssatop\rhup{r} \times \partial_2 \ssatop\rhup{r}\mid$
( $\partial_1$,  $\partial_2$ and $\cdot$ denoting differentiation with
respect to $\varphi^1,~\varphi^2$ and $t$, respectively).
Two cases of special interest are $ a = \sqrt g$ and $a = \sqrt{1-g}$,
the former corresponding to a reduction of the self-dual Yang-Mills
equations in ${\bf R}^4$ with the gauge group SDiff $\Sigma_t$
(see e.g. [9], [10]), while the latter corresponds to the surface
sweeping out a minimal 3-dimensional hypersurface in Minkowski-space
(this was derived in [11]).
In order to show the equivalence of (1) with 3-dimensional steady-state
isentropic irrotational gas-dynamics ( and explicitly solve the case
$a = \sqrt g$) let me write (1) in a slightly different way, namely
$$
\dot x = \gamma \{ y,~z \}~, \qquad \dot y = \gamma \{ z,~x \}~,
\qquad \dot z = \gamma \{ x,~y \}~,
\eqno{(2)}
$$
where $\gamma ={a(g) \over \sqrt{g}}$ and $\{$ .,. $\}$ denotes the
Poisson-bracket
${\partial \over \partial\varphi^1} {\partial \over \partial\varphi^2}-
{\partial \over \partial\varphi^2} {\partial \over \partial\varphi^1}$.
Choosing the unknown functions $x$, $y$ and $z$ as the independent(!)
variables,
$$
t,~\varphi^1,~ \varphi^2 ~~\rightarrow~~
x^1 = x (t,~\varphi^1,~ \varphi^2)~,~~x^2 = y (t,~\varphi^1,~
\varphi^2)~,
x^3 = z (t,~\varphi^1,~ \varphi^2)~
\eqno{(3)}
$$
(possible as long as $\dot{\ssatop\rhup{r}} \neq 0 $), and
$$
B_1 = \gamma \{ y,~z \}, \qquad B_2 = \gamma \{ z,~x \}, \qquad B_3 =
\gamma \{ x,~y \}~,
\eqno{(4)}
$$
expressed in the new coordinates, as the new dependent variables,
such that
$$
\partial_t ~\ssatop{\wedge}{=}~ \mathop{B}^\rhup \cdot \mathop{\nabla
}^\rhup~,
{}~~\gamma  \cdot \{ ~\cdot , ~\mathop{r}^\rhup \} ~\ssatop{\wedge}{=}~
\mathop{B}^\rhup \times \mathop{\nabla }^\rhup~~,
\eqno{(5)}
$$
one finds
$$
\mathop{B}^\rhup \cdot \mathop{\nabla }^\rhup B_i
- B_i \mathop{\nabla }^\rhup \cdot \ssatop\rhup{B}
+ \mathop{B}^\rhup~\partial_i \mathop{B}^\rhup
= B_i ~~ \mathop{B}^\rhup~\cdot~{{\ssatop\rhup{\nabla } \gamma
}\over{\gamma }}~~,
\eqno{(6)}
$$
by taking the time-derivative of (4), using (2) and (5).

These equations for $\ssatop\rhup{B}(\ssatop\rhup{x})$ can be reduced
to a single scalar field equation by simply noting that
$\ssatop\rhup{\nabla }t$
(calculated from (3), using (2)) is actually equal to
$\ssatop\rhup{B}/\ssatop\rhup{B}^2$.
Alternatively, writing
$\ssatop\rhup{B}= \ssatop\rhup{C}/(\ssatop\rhup{C}^2)$
- which is reminiscent of an invariance - transformation of the Lam\'e
system (see e.g. [12] ), and was already used in [10] to linearize
(in an equivalent, spinorial, notation) the SU($\infty $) Nahm
equations -
leaves (6) unchanged, except for a crucial flip of sign in the pure
divergence term,
$$
\mathop{C}^\rhup \cdot \mathop{\nabla }^\rhup C_i - C_i \mathop{\nabla
}^\rhup
\cdot \mathop{C}^\rhup - \mathop{C}^\rhup \partial_i \mathop{C}^\rhup
= C_i \ssatop\rhup{C}~\cdot~\ssatop\rhup{\nabla }~\ln \gamma ~~.
\eqno{(7)}
$$
Multiplying by $C_i$ (and summing over $i=$1, 2, 3) one finds that
$\ssatop\rhup{\nabla }~\cdot~\ssatop\rhup{C}
= - \ssatop\rhup{C}~\cdot~\ssatop\rhup{\nabla } \ln \gamma $.
Hence $\ssatop\rhup{C}~\cdot~\ssatop\rhup{\nabla }C_i =
\ssatop\rhup{C}~\partial_i~\cdot~\ssatop\rhup{C}$, which together with
$$
\ssatop\rhup{C}~\cdot~(\ssatop\rhup{\nabla } \times \ssatop\rhup{C}) =
0
\eqno{(8)}
$$
(this being a consequence of the Jacobi-identity
$\{~\{ x,~y \},~z \} + {\rm cycl.} = 0 $) implies
$$
\ssatop\rhup{C} = \ssatop\rhup{\nabla } f~~.
\eqno{(9)}
$$
The only equation to be solved is therefore
$$
\ssatop\rhup{\nabla } ( \tilde \gamma  ~\ssatop\rhup{\nabla } ~f) =
0~~,
\eqno{(10)}
$$
with $\tilde \gamma $ being a definite function of
$(\ssatop\rhup{\nabla } f)^2$
(invert $\gamma (g)$ to obtain $g(\gamma )$ and solve
$g(\gamma ) \cdot \gamma ^2 \cdot (\ssatop\rhup{\nabla } f)^2$ $= 1$
for
$\gamma  = \tilde \gamma  ((\ssatop\rhup{\nabla } f)^2 )$.

For the case $\tilde \gamma =1$, which before has been solved in rather
different ways and contexts (e.g., known as the `Eden-model'
in the context of growing surfaces, [13]), this means that the time
$t=f(\ssatop\rhup{x})$, at which the surface passes the point
$\ssatop\rhup{x}$, is a harmonic function of the three space variables
$x$, $y$, and $z$.
This observation provides a very intuitive understanding of the
solution(s)  of the corresponding equation(s) (1).
To illustrate this by an example, let
$$
t (x,~y,~z) = z^2 - x^2~~.
\eqno{(11)}
$$
This solution of Laplace's equation corresponds to two sheets of
hyperboloids in the $x-z$ plane (infinitely extending perpendicular to
this plane) which move towards the $y$-axis,
becoming `singular' at $t=0$, and then moving away from the $y$-axis,
in directions perpendicular to the incoming ones.
In this case every $\ssatop\rhup{x} \in {\bf R}^3$ is passed exactly
once.
Generally, any $t(\ssatop\rhup{x})$ that is finite for all
$\ssatop\rhup{x} \in {\bf R}^3$ will be a superposition of harmonic,
homogeneous polynomials, of course.

Going back to the general case ($\tilde \gamma  \neq {\rm const.}$),
one notes that eq. (10) means that
all equations of type (1) can (at least locally) be written as
Lagrangian equations of motion for a scalar field $f(\ssatop\rhup{x})$
(which is the time at which the surface passes the point
$\ssatop\rhup{x}$), with the Lagrangian density $\cal{L}$ being
${1 \over 2}$ the integral of $\gamma $ (expressed as a function of
$(\ssatop\rhup{\nabla }f)^2)$.
So the following interesting correction with fluid dynamics emerges:
viewing $\ssatop\rhup{\nabla }f = \ssatop\rhup{V}$ as the velocity,
and $\tilde \gamma  \not\equiv {\rm cont}$ as the mass density $\rho $
of an
irrotational inviscid 3 dimensional gas, (10) is the continuity
equation
for time independent (`steady state') flows, while the Euler equation,
$$
\ssatop\rhup{\nabla } \dot f + \ssatop\rhup{\nabla } f \cdot
\ssatop\rhup{\nabla }
(\ssatop\rhup{\nabla } f ) + {1 \over{\rho }} \ssatop\rhup{\nabla }
(P(\rho )) = 0~~,
\eqno{(12)}
$$
(for the steady state case, $\ssatop\rhup{\nabla } \dot f = 0$)
determines
$\rho =\rho $ ($(\ssatop\rhup{\nabla }f)^2 =w$) in exactly the way
needed for
a consistent interpretation of $\tilde \gamma $ as the mass density,
provided the pressure $P(\rho )$ is chosen such that
$$
{{dP(\rho (w))}\over{dw}} = - {1 \over 2} \rho (w)~~.
\eqno{(13)}
$$
For
$$
\rho  = ( (\ssatop\rhup{\nabla }f)^2 - 1)^{-1/2}~~,
\eqno{(14)}
$$
which corresponds to $a = \sqrt{1-g}$ in eq. (1) (i.e. the relativistic
minimal hypersurface case), the equations of motion, (10), read
$$
\ssatop\rhup{\nabla }^2 f = {1 \over 2}~
{{\ssatop\rhup{\nabla }f \cdot \ssatop\rhup{\nabla
}((\ssatop\rhup{\nabla }f)^2)}\over%
{(\ssatop\rhup{\nabla }f)^2 -1}}~~,
\eqno{(15)}
$$
corresponding to the Lagrangian
$$
{\cal L} = \sqrt{(\ssatop\rhup{\nabla } f )^{2} - 1 }~~.
\eqno{(16)}
$$
It is interesting to note that (13) yields the
K\'arm\'an-Tsien-Chaplygin
equation of state,
$$
P(\rho ) = - {1 \over \rho }~~,
\eqno{(17)}
$$
as long as
$$
\rho  = ( (\ssatop\rhup{\nabla }f)^2 + \epsilon  )^{-{1\over 2}}~~,
\eqno{(18)}
$$
for \undertext{all} $\epsilon $.
This means that the relativistic ($\epsilon =-1$) \undertext{and} the
Euclidean
($\epsilon =+1$) minimal hypersurface problem, as well as the
minimization of
$\int d^3 x \mid \ssatop\rhup{\nabla }f \mid $ ($\epsilon =0$) are all
related to
K\'arm\'an-Tsien irrotational gas-dynamics, - the first corresponding
to the supersonic regime ($\ssatop\rhup{V}^2 = (\ssatop\rhup{\nabla
}f)^2 >
{{dP}\over{d\rho }} \equiv ({\rm velocity~~of~~sound})^2 \equiv c^2$),
the second to the subsonic regime,
and the third ($\epsilon =0$) corresponding to the case of the
Mach-number
M: $= \mid \ssatop\rhup{V}\mid/c$ being exactly equal to one.

However, as it was demonstrated in [15] that the relativistic minimal
hypersurface-problem also corresponds to a time\undertext{dependent}
K\'arm\'an-Tsien gas in one space-dimension lower, the rather unique
fact emerges that
a d-dimensional steady state K\'arm\'an-Tsien gas
is equivalent to a time-dependent d-1 dimensional one.
Of course, this has to do with the hidden relativistic invariance, and
is therefore, a posteriori, not surprising (and can easily be checked
directly).

Finally, let me also derive (10) by first obtaining a single
second-order
equation for $z(t,~x,~y)$ from (1), and then interchanging the role of
$z$ and $t$ as dependent, resp. independent, variable:
So, consider first the transformation
$$
\varphi^0 = t,~\varphi^1,~\varphi^2 ~~\rightarrow~~
y^0 = t,~~~y^1 = x(t,~\varphi^1,~\varphi^2),~
y^2 = y(t,~\varphi^1,~\varphi^2)
\eqno{(19)}
$$
for (1), which allows one to derive from (2) a pair of first order
differential equations for $J=\{ x,~y \}$ and $z$
(both viewed as functions of the new variables; $\gamma =\gamma (g)$
with
$g = J^2 \cdot (1 + (\ssatop\rhup{{\mib \nabla }}z)^2)$,
$\ssatop\rhup{{\mib \nabla }} = \Big( {\partial\over{\partial y^1}},~
{\partial\over{\partial y^2}} \Big)^{tr}$),
$$
\dot z = \gamma J(1+ (\ssatop\rhup{{\mib \nabla }}z)^2 )~,~~~
\dot J = - J^2 \ssatop\rhup{{\mib \nabla }}(\gamma \ssatop\rhup{{\mib
\nabla }}z )
\eqno{(20)}
$$
(cp. [11]). One can then verify that the corresponding second order
equation for $z$,
$$
\ssatop\rhup{{\mib \nabla }} ( \tilde \gamma  \ssatop\rhup{{\mib \nabla
}}z) =
\ssatop\bullet{
\Big( \tilde \gamma  {{(1+(\ssatop\rhup{{\mib \nabla }}z)^2)}\over{\dot
z}} \Big)}~~,
\eqno{(21)}
$$
follows from the Lagrangian
$$
\tilde{\cal L} = \dot z {\cal L}
\Big( w = {{1+(\ssatop\rhup{{\mib \nabla }}z)^2}\over{{\dot z}^2}}
\Big)
\eqno{(22)}
$$
if $2{{\partial {\cal L}}\over{\partial w}} = \tilde \gamma  (w)$
(the latter being defined by $\tilde \gamma  = \gamma $ ($g =
{{1}\over{\tilde \gamma ^2 \cdot
w}}$ ) ).
Note that (21) remains nonlinear if $\tilde \gamma  = {\rm const}$,
in contrast with (10), while the transformation
$$
x_1,~x_2,~x_3 ~~\rightarrow~~ y_0 = f(\ssatop\rhup{x}), ~~y_1 = x_1,
{}~~y_2 = x_2
\eqno{(23)}
$$
(resp. $y_0~y_1~y_2 \rightarrow x_1=y_1$, $x_2=y_2$,
$x_3=z(y_0, y_1, y_2)$ ) provides a one to one correspondence between
$$
S = \int d^3 y~ \dot z \cdot {\cal L}
\Big(
{{(\ssatop\rhup{{\mib \nabla }}z)^2 + 1}\over{{\dot z}^2}}
\Big)
\eqno{(24)}
$$
and
$$
S = \int d^3 x {\cal L} ( (\ssatop\rhup{\nabla } f)^2 )~~,
\eqno{(25)}
$$
hence confirming $f(\ssatop\rhup{x})$ as the time at which the
surface passes $\ssatop\rhup{x}=(x_1,~x_2,~x_3)$.

\vskip7mm
\noindent
{\bf Acknowledgement}

I would like to thank A. Anderson, A. Bray, J. Chapman, D. Giulini,
G. Horowitz, M. Kardar, C. Klimcik, J. Krug, L. Mason, P. McCarthy and
B. Palmer, as well as T. Inami and K. Takasaki, for valuable
discussions, the Isaac Newton Institute for hospitality,
and the Deutsche Forschungsgemeinschaft for
financial support.

\vskip7mm
\noindent
{\bf Addendum:}

Contemplating the possibility that special choices of $\gamma $ may
correspond to integrable 3-manifolds that are connected to the
SU($\infty $)
Nahm case by generalized B\"acklund-transformations, it might be
useful to note various `zero-curvature or Lax-type' representations
for the case of constant $\gamma $, and axially symmetric surfaces.
Firstly, (if $\gamma =1$) (21) can be written in Lax-form (with
spectral
parameter, $\lambda $),
$$
\dot L = [ L,~M]~~,
\eqno{(26)}
$$
by defining
$$
\eqalign{
L(\lambda ) & = {{\dot z}\over{1+\partial z \bar\partial z}}( (
\partial z + {i \over \lambda }) \bar\partial - (\bar\partial z +
i\lambda )\partial)\cr
M(\lambda ) & = {{\dot z}\over{1+\partial z \bar\partial z}}( i\lambda
\partial - (\partial z)\bar\partial )~~,\cr
}
\eqno{(27)}
$$
where $\partial : = \partial_x - i\partial_y$,
$\bar\partial = \partial_x + i\partial_y$.
One may also just take
$$
\eqalign{
L & = {{\dot z}\over{1+\partial z \bar\partial z}} \bar\partial \cr
M & = {{- \dot z}\over{1+\partial z \bar\partial z}} (\partial z)
\bar\partial~~.\cr
}
\eqno{(28)}
$$
Actually, any equation of motion of the form
$$
{d \over dt}~F(x,~y;~z,~\partial z,~\bar\partial z,~ \dot z, \cdots) =
\ssatop\rhup{\nabla }^2 z
\eqno{(29)}
$$
is representable in Lax-form by letting
$$
L = {1 \over F} \bar\partial~, \qquad M = {-1 \over F} (\partial z)
\bar\partial~~.
\eqno{(30)}
$$
To obtain conserved quantities for the SU($\infty $)-Nahm equations
(with compact $\Sigma_t$) it is easiest to represent the original
equations, (2), in Lax-form on the Poisson-Lie algebra of functions
(on $\Sigma_t$), rather than vectorfields,
$$
\eqalign{
{\cal L} & = {1 \over \lambda } (x + iy) - 2iz + \lambda  (x-iy) \cr
{\cal M} & = iz - \lambda (x-iy) \cr
\dot{\cal L} & = \{ {\cal L},~{\cal M} \} \cr
}
\eqno{(31)}
$$
which implies that
$$
Q_{lm} : =
\Big( {\partial^m \over \partial\lambda ^m}
\int_{\Sigma_t} d\varphi^1 d\varphi^2 ~{\cal L}(\lambda )^l \Big)
\mid_{\lambda =0}
\eqno{(32)}
$$
is time-independent; the conserved densities are just the harmonic
homogenous polynomials (note that $(\ssatop\rhup{\nabla }{\cal L})^2
=0$).
(10)$_{\gamma =1}$ on the other hand, is equivalent to
$$
[ ( L_1 + iL_2 ) + \lambda (L-iL_3), ~~(L + iL_3) - \lambda (L_1 -
iL_2)] = 0~~,
\eqno{(33)}
$$
$$
L: = {{\ssatop\rhup{\nabla }f}\over{(\ssatop\rhup{\nabla }f)^2}} \cdot
\ssatop\rhup{\nabla }~,
{}~~~~\ssatop\rhup{L} : = {{\ssatop\rhup{\nabla
}f}\over{(\ssatop\rhup{\nabla }f)^2}}
\times \ssatop\rhup{\nabla }~~,
\eqno{(34)}
$$
due to the identity
$$
[ L,~L_i ] + {1 \over 2} \epsilon _{ijk} [ L_j,~ L_k ] =
{{- \ssatop\rhup{\nabla }^2 f}\over{(\ssatop\rhup{\nabla }f)^2}} L_i~~.
\eqno{(35)}
$$
This leads to the possibility of representing axially symmetric
surface motions by
$$
[ \gamma ^{-1} ( L_1 + iL_2),~L + iL_3 ] =0~~,
\eqno{(36)}
$$
as $[\gamma ,~L_3] = 0$ if $f=f(\sqrt{x^2+y^2},~z)$.

\endpage

\noindent
{\bf Reference}

\item{[1]}P. Pelce' ; 'Dynamics of Curved Fronts' Academic Press,
New York 1988.

\item{[2]}K.A. Brakke ; `The Motion of a Surface by its Mean Curvature'
Princeton University Press, Princeton, NJ 1978.

\item{[3]}S. Osher, J.A. Sethian ; J. of Computational Physics {\bf 79}
(1988) 12.

\item{[4]}L.C. Evans, J. Spruck ; J. of Diff. Geom. {\bf 33} (1991)
635.
Y.G. Chen, Y. Giga, S. Goto; J. Diff. Geom. {\bf 33} (1991) 749.

\item{[5]}M. Struwe ; `Geometric Evolution Problems', Lecture Notes
( to appear in the Park City Geometry Series of the AMS )

\item{[6]}K. Nakayama, M. Wadati ; J. of the Phys.Soc. of Japan Vol. 62
(1993) 1895.

\item{[7]}R.I. McLachlan, H. Segur; `A Note on the Motion of Surfaces'
Boulder preprint PAM \#{\bf 162} ( 1993 ).

\item{[8]}A.I. Bobenko; `Surfaces in Terms of $2\times 2$ Matrices.
Old and New Integrable Cases' SFB288 preprint \#{\bf 66} (Berlin) 1993.

\item{[9]}E.G. Floratos, G.K. Leontaris; Phys. Lett. B {\bf 223}
(1989) 153.

\item{[10]}R.S. Ward; Phys. Lett. B {\bf 234} (1990) 81.

\item{[11]}M. Bordemann, J. Hoppe; Phys. Lett. B {\bf 325} (1994) 359.

\item{[12]}B.G. Konopelchenko, W.K. Schief ; `Lame' and
Zakharov-Manakov Systems : \break
Combescure, Darboux and Baecklund Transformations' 1993.

\item{[13]}M. Kardar; private communication.

\item{[14]}A. Bray, J. Krug; private communication.

\item{[15]}M.Bordemann, J.Hoppe; Phys.Lett. B {\bf 317} (1993) 315.

\bye